\documentclass[letterpaper, 10 pt, conference]{ieeeconf}  

\IEEEoverridecommandlockouts                              

\usepackage{cite}
\usepackage{amsmath,amssymb,amsfonts}
\usepackage{algorithmic}
\usepackage{graphicx}
\usepackage{textcomp}
\usepackage{setspace}

\usepackage{graphics}
\usepackage{epsfig}
\usepackage{times} 
 \usepackage[dvipsnames]{xcolor}
\usepackage{xurl}
\usepackage{hyperref}
\usepackage{cleveref}

\usepackage{bm}
\usepackage{breqn}
\usepackage{amsmath}
\usepackage{amssymb}
\usepackage{tabularx}

\usepackage{amsthm}

\newcommand{\integers}{\mathbb{Z}}
\newcommand{\naturals}{\mathbb{N}}
\newcommand{\reals}{\mathbb{R}}
\newcommand{\R}{\reals}
\newcommand{\Rnonneg}{\reals_{\geq 0}}
\newcommand{\Rplus}{\reals_{>0}}

\renewcommand{\S}{\mathbb{S}}

\newcommand{\pd}{\S_{++}}

\newcommand{\E}{\mathbb{E}}

\newcommand{\GP}{\mathsf{GP}}
\renewcommand{\k}{\mathsf k}
\newcommand{\kt}{\mathsf {k_T}}
\newcommand{\ks}{\mathsf {k_S}}
\newcommand{\K}{\mathsf{K}}

\newcommand{\Dcal}{\mathcal{D}}

\newcommand{\Fcal}{\mathcal{F}}

\newcommand{\Ncal}{\mathcal{N}}
\newcommand{\Ocal}{\mathcal{O}}

\newcommand{\Ucal}{\mathcal{U}}

\newcommand{\Xcal}{\mathcal{X}}

\renewcommand{\sf}[1]{\mathsf{#1}}

\newcommand{\eqn}[1]{\begin{align} #1 \end{align}}
\newcommand{\eqnN}[1]{\begin{align*} #1 \end{align*}}

\newcommand{\bmat}[1]{\begin{bmatrix}#1\end{bmatrix}}

\newcommand{\norm}[1]{\left\Vert #1 \right \Vert}
\newcommand{\abs}[1]{\left | #1 \right |}

\newcommand{\argmin}[1]{\underset{#1}{\text{argmin}}}

\newcommand{\inner}[1]{\left \langle #1 \right \rangle}

\theoremstyle{plain}
\newtheorem{theorem}{Theorem}
\newtheorem{corollary}[theorem]{Corollary}
\newtheorem{lemma}[theorem]{Lemma}

\theoremstyle{definition}
\newtheorem{definition}{Definition}
\newtheorem{assumption}{Assumption}
\newtheorem{problem}{Problem}
\newtheorem{remark}{Remark}

\theoremstyle{remark}

\newtheorem{example}{Example}

\crefname{assumption}{assumption}{assumptions}
\Crefname{assumption}{Assumption}{Assumptions}

\crefname{problem}{problem}{problems}
\Crefname{problem}{Problem}{Problems}

\usepackage{acronym}
\acrodef{NGPKF}[NGPKF]{Numerical Gaussian Process Kalman Filter}
\acrodef{GP}[GP]{Gaussian Process}
\acrodef{KF}[KF]{Kalman Filter}
\acrodef{DCT}[DCT]{Discrete Cosine Transform}
\acrodef{SDE}[SDE]{Stochastic Differential Equation}
\acrodef{TSD}[TSD]{Target Spatial Distribution}

\usepackage[absolute]{textpos}
\title{\LARGE \bf
Multi-Agent Clarity-Aware Dynamic Coverage with Gaussian Processes
}

\author{Devansh R. Agrawal and Dimitra Panagou
\thanks{*The authors would like to acknowledge the support of the National Science Foundation (NSF) under grant no. 1942907 and grant no. 2223845.}
\thanks{Devansh R Agrawal is with the Department of Aerospace Engineering, University of Michigan, Ann Arbor, USA.
        {\tt\small devansh@umich.edu}.}%
\thanks{Dimitra Panagou is with the Department of Robotics and the Department of Aerospace Engineering, University of Michigan, Ann Arbor, USA. {\tt\small dpanagou@umich.edu}.}%
}

\begin{document}

\begin{textblock*}{20cm}(0.5cm,0.5cm)
Author's copy. To appear at IEEE CDC 2024. 
\end{textblock*}

\maketitle
\thispagestyle{empty}
\pagestyle{empty}

\begin{abstract}
This paper presents two algorithms for multi-agent dynamic coverage in spatiotemporal environments, where the coverage algorithms are informed by the method of data assimilation. In particular, we show that by explicitly modeling the environment using a \acf{GP} model, and considering the sensing capabilities and the dynamics of a team of robots, we can design an estimation algorithm and multi-agent coverage controller that explores and estimates the state of the spatiotemporal environment. The uncertainty of the estimate is quantified using clarity, an information-theoretic metric, where higher clarity corresponds to lower uncertainty. By exploiting the relationship between \acp{GP} and \acfp{SDE} we quantify the increase in clarity of the estimated state at any position due to a measurement taken from any other position. We use this relationship to design two new coverage controllers, both of which scale well with the number of agents exploring the domain, assuming the robots can share the map of the clarity over the spatial domain via communication. We demonstrate the algorithms through a realistic simulation of a team of robots collecting wind data over a region in Austria. 
\end{abstract}

Code and open-sourced Julia packages are available at~\cite{repo}. 

\section{Introduction}

A standard robotic mission is the collection of information that varies both in time and space over a domain of interest. To collect such information optimally, a (team of) robot(s) must reason about the currently available information, the target level of confidence in the information sought, the spatiotemporal evolution of the underlying information, and the robot's sensing capabilities, and (in the case of a team) coordinate the actions of each robot. 

The design of informative path planners and dynamic coverage controllers has long been of interest~\cite{liang2014survey, oktug20083d, bentz2020dynamic}, with a variety of techniques proposed including Voronoi partitioning~\cite{cortes2004coverage}, sampling approaches~\cite{moon2022tigris, bry2011rapidly}, grid/graph based approaches~\cite{cao2013multi, xiao2022nonmyopic} and ergodic search~\cite{mathew2011metrics, dressel2019tutorial}.

In this paper, we define the informative path planning or coverage control problem as follows: we have a team of robots that, at a fixed sampling frequency, measure the spatiotemporal environment at their respective positions. Using these measurements, we update our estimate of the state of the environment (referred to as information assimilation), while simultaneously controlling the robots position to determine the next location from which a measurement should be taken (referred to as the coverage controller). As such, the goal is to design a controller and information assimilation algorithm that efficiently reduce the uncertainty of the estimate of the state of the environment.

We quantify the uncertainty of a stochastic variable using an information-theoretic metric \emph{clarity}, introduced in~\cite{agrawal2023sensor}.  In particular, as the uncertainty of the stochastic variable decreases, (i.e., its differential entropy approaches $-\infty$), the clarity of the random variable approaches~1. Similarly, as the uncertainty increases, the clarity approaches zero.

We model the environment as a spatiotemporal field $f(t, p)$, i.e., a scalar function that varies in time and space; as an example, if the goal is to estimate the windspeed over a spatial and temporal domain, $f(t, p)$ represents the windspeed at any given time $t$ and position $p$. The estimate is a function $\hat f(t, p)$ for each $t, p$, with an associated clarity $q(t, p)$ at each $t, p$. Numerically the state of the environment is a vector representing $\hat f(t, p)$ at a set of grid points. By taking (noisy) measurements of $f$ using the robots at their respective locations, we can improve our estimate $\hat f$ and increase its clarity (i.e., reduce the uncertainty). At the same time, due to the time-varying nature of $f$, the clarity of $\hat f$ decreases for all points not being measured. This balance of information gain and decay will be an important element in designing the algorithms.

A key limitation of many of the methods listed above is that simplified heuristics are used to motivate the cost functions used in the informative path planners. For example, the ergodic search approaches assume that a \ac{TSD} (defined as the desired percentage of time that the robot should spend at any position in the domain) is provided by the user. However, there has been less work on how one can obtain such a target distribution in a principled manner taking into account the sensing capabilities of the robot or the temporal evolution of the state of the environment.

The goal of this paper is to demonstrate how the cost function in informative path planning can be designed in a principled manner based on the assumed model of the environment. In particular, when estimating a spatiotemporal field, a common practice is to model it as a realization of a \ac{GP}~\cite{williams2006gaussian}, and use the robot's measurements to update the estimate of the state of environment. 

Here, we use the connection between \acp{GP} and \acp{SDE}~\cite{sarkka2012infinite, kuper2022numerical, carron2016machine} to analyze the information-gathering capabilities of the robots: given the robot's take measurements at their respective locations, how much does the uncertainty in our estimate of state of the environment reduce? We answer this by quantifying a robot's sensing function and the environment's information decay function. For a point $p$ in the domain, the sensing function defines the rate of increase of clarity at $p$ due to measurements from a robot at position $r$. The decay function quantifies the rate of decrease of clarity due to the time-varying nature of $f(t, p)$.  We use these functions to design coverage controllers that respect the rate of change of clarity when designing trajectories. 

This paper has three main contributions: (A)~We use clarity~\cite{agrawal2023sensor} to quantify the rate of change of uncertainty at a position~$p$ due to measurements made by a robot at a (possibly different) position~$r$. Integrated over the mission domain, this quantifies the value of the robot being at position $r$. (B)~We use this relation to propose two coverage controllers. (C)~Being feedback controllers, we show how they scale naturally to the multi-agent setting. Finally, we demonstrate the algorithms using a realistic simulation, where a team of aerial robots explore a region of Austria, and estimate the wind speed over this region.

The two coverage algorithms proposed bear resemblance to the controllers in~\cite{bentz2020dynamic} and~\cite{naveed2024eclares}. The first, referred to as the \emph{direct controller}, directly chooses a control input to maximize the clarity of the state of the environment. The second, referred to as the \emph{indirect controller}, computes a \ac{TSD} based on the time required to increase the clarity to a given target value.


\section{Preliminaries}

\subsubsection*{Notation}
$\integers$ is the set of integers, $\naturals = \{ 0, 1, 2, ... \}$ is the set of naturals. $\R, \Rnonneg, \Rplus$ denote the sets of reals, nonnegative reals, and positive reals. $I_N$ denotes the $N$-dimensional identity matrix. $\pd^n$ denotes the set of symmetric positive definite matrices in $\R^{n \times n}$. For $A \in \pd^n$, $\sqrt{A} \in \pd^n$ is the unique matrix such that $\sqrt{A} \sqrt{A} = A$. For $v \in \R^n$, the $i$-th entry is $[v]_i$. Similarly, for $M \in \R^{N \times M}$, the $(i, j)$-th entry is $[M]_{(i, j)}$. $A \otimes B$ denotes the Kronecker product of $A, B$.

We consider a problem with $N_R$ robots, exploring a $d$-dimensional domain $\Dcal \subset \R^d$. Each robot has a state in $\Xcal \subset \R^n$, $n \geq d$. The state of the environment will be represented numerically at a set of $N_G$ grid points.

\subsection{Clarity}



The information metric \emph{clarity} was introduced in~\cite{agrawal2023sensor} and is based on differential entropy:
\begin{definition}\cite[Ch. 8]{thomas2006elements}
The \emph{differential entropy} $h[X] \in (-\infty, \infty)$ of a continuous random variable $X$ with support $S$ and density $\rho: S \to \R$ is 
\eqn{
h[X] = - \int_{S} \rho(x) \log \rho(x) dx. \label{eqn:diff_entropy}
}
\end{definition}

Notice that as the uncertainty in $X$ decreases, the entropy approaches $h[X] \to -\infty$. Clarity is defined in terms of differential entropy. 
\begin{definition}
Let $X$ be a $n$-dimensional continuous random variable with differential entropy $h[X]$. The \emph{clarity} $q[X] \in (0, 1)$ of $X$ is defined as: 
\eqn{
q[X] = \left(1+\frac{\exp{(2 h[X])}}{(2 \pi e)^n}\right)^{-1}. \label{eqn:clarity}
}
\end{definition}

In other words, the clarity $q[X]$ about a random variable $X$ lies in $(0, 1)$, where $q \to 0$ corresponds to the case where the uncertainty in $X$ is infinite, and if $X$ is perfectly known in an idealized (noise-free) setting,  $q[X] = 1$. For a scalar Gaussian random variable $X \sim \Ncal(\mu, \sigma^2)$, the clarity is $q[X] = 1 / (1 + \sigma^2)$. 

In an estimation context, we use clarity to quantify the quality of our estimate: as the clarity increases towards 1, the uncertainty of our estimate decreases towards 0. In~\cite{agrawal2023sensor} it was shown that when $X$ is estimated using a Kalman filter, the clarity dynamics of the estimate of $X$ can be obtained in closed form. 

\subsection{Gaussian Processes}

A \ac{GP}~\cite[Ch. 2]{williams2006gaussian} is a (scalar) stochastic process that is fully defined by the mean function $\mathsf m : \Dcal \to \R$ and a kernel $\k : \Dcal \times \Dcal \to \R$:
\eqn{
f(p) \sim \GP(\mathsf m(p), \k(p, p')), 
}
where $m$ and $\k$ are defined as
\begin{subequations}
\eqn{
\mathsf{m}(p) &= E[f(p)],\\
\k(p, p') &= E[ (f(p) - \mathsf m(p)) ( f(p') - \mathsf m(p'))].
}
\end{subequations}

Given a set of $N$ measurements $\{ y_k \}_{k=1}^{N}$ taken at positions $\{ p_k \}_{k=1}^{N}$, we can update our posterior estimate of $f$, as described in~\cite[Ch. 2]{williams2006gaussian}. 

For two set of points $P_A = \{ a_i \}_{i=1}^{N}$ and $P_B = \{ b_i \}_{i=1}^{M}$, the kernel matrix $\K_{AB} \in \R^{N \times M }$ is the matrix such that  $[\K_{AB}]_{(i, j)} = \k(a_{i}, b_j)$.  

\subsection{Spatiotemporal Gaussian Processes}
\label{section:ngpkf}
\label{section:sde}

The goal is to estimate a spatiotemporal field, i.e., to estimate a function $f(t, p)$,  $f: \R \times \Dcal \to \R$ using measurements obtained by robots.\footnote{For simplicity of exposition, we assume the spatiotemporal field has scalar outputs. For multidimensional outputs, we repeat for each dimension independently.} Here $t \in \R$ denotes time, and $\Dcal \subset \R^d$ is spatial domain of interest. The measurements (defined in~\eqref{eqn:robot_measurements}) are noisy measurements of $f$ at a fixed sampling period from each robot's position at the sampling time.

While a standard \ac{GP} can directly handle the spatiotemporal case, we can achieve significant computational efficiency by explicitly separating the spatial and temporal dimensions and exploiting the equivalence between spatiotemporal \acp{GP} and \acp{SDE}. Effectively, we can convert a Bayesian inference problem into a Kalman Filtering problem, thereby reducing memory and computational cost. We assume the following:
\begin{assumption}
\label{assumption:separable}
Suppose the spatiotemporal field $f: \R \times \Dcal \to \R$ is a realization of a zero-mean \ac{GP}:
\eqn{
f(t, p) &\sim \GP(0, \k(t, p, t', p')),\\
\k(t, p, t', p') &= \kt(t, t') \ks(p, p'),
}
where the kernel is separable in space and time, and the temporal kernel is isotropic, i.e., $\kt(t, t')$ only depends on $\abs{t' - t}$. 
\end{assumption}

Under~\Cref{assumption:separable}, it is known that realizations of a \ac{GP} are also realizations of a \ac{SDE}~\cite{carron2016machine}. This fact is derived through the Wiener-Khinchin theorem~\cite[Ch. 12]{sarkka2019applied}, and in the interest of space, the readers are referred to \cite{carron2016machine} or~\cite[Appendix]{agrawal2024multi} for full derivations. 

The key idea is that if $h(t) \sim \GP(0, \kt(t,t'))$ is a realization of a (temporal) \ac{GP}, it is equal to the output of a transfer function applied to a realization of a white noise process. By expressing the transfer function in state-space form, we arrive at a \ac{SDE} such that a realization of the \ac{SDE} is equal to $h$. 

In the spatiotemporal case, let $P_G = \{ p_i \}_{i=1}^{N_G} \subset \Dcal$ be a set of $N_G$ (possibly non-uniform) grid points over the spatial domain. Let $\bm f(t) \in \R^{N_G}$ be a vector such that the $i$-th entry is the value of the spatiotemporal field at the $i$-th grid point, $[\bm f(t) ]_{i} = f(t, p_i)$. Then, the \ac{SDE} for the system comprises of $N_G$ independent stochastic processes~\eqref{eqn:SDE_Si}, that get spatially correlated based on the spatial kernel~\eqref{eqn:SDE_f}. Mathematically, 
\begin{subequations}
    \label{eqn:SDE}
\eqn{
& 
\begin{cases}
    ds_i(t) = A s_i(t) dt + B dW_i(t)\\
    z_i(t) = C s_i(t)\\
    s_i(0) \sim \Ncal(0, \Sigma)
\end{cases}, \label{eqn:SDE_Si}\\
&\bm f(t) = \sqrt{\sf K_{GG}} \bm z(t) = \sqrt{ \sf K_{GG}} (I_{N_G} \otimes C) \bm s, \label{eqn:SDE_f}
}
\end{subequations}
Here $s_i(t) \in \R^{n_k}$ is a state at each grid point.\footnote{$n_k$ depends on the temporal kernel. For the Matern 1/2, 3/2, and 5/2 kernels, $n_k = 1, 2, 3$ respectively.} $\bm s = \bmat{ s_1^T & \cdots & s_G^T }^T \in \R^{n_k N_G}$ is a stacked vector representing the state of the entire environment; $\bm f(t) = \bmat{f(t, p_1) & \cdots & f(t, p_G)}^T \in \R^{N_G}$ is a stacked vector comprising the value of the field at each grid point;  $W_i$ is a standard Wiener process, independent for each grid point. The matrices $A \in \R^{n_k \times n_k}$, $B \in \R^{n_k \times 1}$, $C \in \R^{n_k \times l}$ are constant matrices that only depend on the temporal kernel $\kt$. $\Sigma \in \pd^{n_k}$ is the matrix that solves $A\Sigma + \Sigma A^T = - BB^T$. $\sf K_{GG} \in \pd^{G}$ is the spatial kernel matrix, i.e.,  $[\sf K_{GG}]_{ij} = \ks(p_i, p_j)$.


\begin{example}
\label{example:1}
The Matern-1/2 temporal kernel is $\kt(t, t') = \sigma_t^2 \exp{\left(-\lambda_t \abs{t-t'}\right)}$ for hyperparameters $\lambda_t, \sigma_t > 0$. The state-space model has dimension $n_k = 1$, and matrices $A = \bmat{-\lambda_t}$, $B = \bmat{1}$, $C = \bmat{\sqrt{2\lambda_t} \sigma_t}$. Derivations and expressions for Matern-3/2 and Matern-5/2 kernels can be found in~\cite[Appendix]{agrawal2024multi}.
\end{example}

\subsection{Ergodic Control}
\label{sec:ergodic}

Ergodic control~\cite{mathew2011metrics, dressel2019tutorial} is a technique to generate robot trajectories that cover a domain  $\Dcal = [0, L_1] \times \cdots \times [0, L_d] \subset \R^d$, such that the trajectories have a spatial (position) distribution that closely matches a specified \ac{TSD}, as explained below.

The \ac{TSD} is a function $\phi: \Dcal \to \R$ such that the $\phi(p)$ denotes the desired time a robot should spend at position $p$. Given a robot's (position) trajectory $\xi  : [0, T] \to \Dcal$ the trajectory's spatial distribution is defined as $c_\xi : \Dcal \to \R$, where for any $p \in \Dcal$ 
\begin{equation}
    c_{\xi}(p) = \frac{1}{T}\int_{0}^{T} \delta(p - \xi(\tau))) d\tau.
\end{equation}
Here $\delta: \R^d \to \R$ is the Dirac delta function. 

The \emph{ergodicity} $E > 0$ of a trajectory $\xi$ measures the difference between the robot trajectory's spatial distribution and the target spatial distribution: 
\eqn{
E = \norm{ c_{\xi} - \phi}_{H^{-s}}^2
}
where $\norm{\cdot}_{H^{-s}}$ is the Sobolev space norm of order $s = (d+1)/2$, defined in~\cite{mathew2011metrics}:
\eqn{
\label{eqn:sobolev}
\norm{c_\xi - \phi}_{H^{-s}}^2 &= \sum_{l \in \naturals^d} \Lambda_l (\hat c_l - \hat \phi_l)^2
}
where $\Lambda_l \in \R$ is a weighting coefficient, and $\hat{(\cdot)}_l$ is the $l$-th element of the \ac{DCT} of the function $(\cdot)$, e.g.
\eqn{
\hat \phi_l = \langle \phi_l, b_l \rangle = \int_{p \in \Dcal} \phi_l(p) b_l(p) dp
}
where $b_l: \Dcal \to \R$ is the $l$-th basis function. We refer the reader to~\cite{mathew2011metrics} for further details.

$E$ is a function-space norm measuring the difference between the TSD and the spatial distribution of the trajectory. The key benefit of the Sobolev norm is that it prioritizes matching the low spatial frequency differences between $c$ and $\phi$ before matching the high spatial frequencies. This means that the controllers have a multiscale-spectral nature, where they prioritize covering the domain globally, before returning to the gaps and covering them~\cite{mathew2011metrics}.

In~\cite{mathew2011metrics} a feedback controller is derived for single and double-integrator robot models that minimizes the ergodicity. Various extensions have been presented in, for example,~\cite{dressel2019tutorial, dong2023time} to address other robot models and other goals. 

\section{Problem Statement}


Consider a team of $N_R > 0$ robots, each with dynamics 
\eqn{
\label{eqn:robot_dynamics}
\dot x_i = F(x_i) + G(x_i) u_i,
}
where $x_i \in \Xcal \subset \R^n$ is the $i$-th robot's state, and $u_i \in \Ucal \subset \R^m$ is its control input. The position of each robot is $r_i = \Phi(x_i) \in \R^d$, i.e., $\Phi : \Xcal \to \Dcal$ extracts the position.

Each robot makes measurements of the spatiotemporal field at a fixed sampling period $\Delta T > 0$, 
\begin{subequations}
\label{eqn:robot_measurements}
\eqn{
y_{k, i} &= f(t_k, \Phi(x_i(t_k))) + w_{k, i},\\
w_{k, i} &\sim \Ncal(0, \sigma_m^2), 
}
\end{subequations}
that is, $y_{k, i} \in \R$ is a scalar measurement output by the $i$-th robot at the $k$-th timestep, $t_k = k \Delta T$. Each measurement is perturbed by zero-mean Gaussian noise with standard deviation $\sigma_m$.

We assume each robot determines its control inputs, but that the information from each robot is assimilated centrally. We assume the robots can always communicate with the central agent, sending the measurements and receiving a map of the current clarity at each $p \in \Dcal$. 

\begin{problem}
\label{problem:main}
    Consider a team of $N_R > 0$ robots, each with dynamics~\eqref{eqn:robot_dynamics} and measurements~\eqref{eqn:robot_measurements}, exploring a domain $\Dcal$. Let $f : \R \times \Dcal \to \R$ be a spatiotemporal field to be estimated satisfying~\Cref{assumption:separable}. Design a coverage control algorithm for each robot, and an estimation algorithm to fuse measurements $y_k$ into an estimate of $f$.
\end{problem}
The mathematical form of the coverage objective is delayed until~\Cref{sec:coverage_controllers}. The estimator will be the optimal estimator in a least-squares sense, discussed in~\Cref{sec:info_assimilation}. 

In addressing \Cref{problem:main}, we address two questions: (A)~how does the information assimilation algorithm inform the value of taking measurements at a robot position $x \in \Dcal$ on the quality of information at a different position $p \in \Dcal$, and (B)~how should one design coverage controllers to exploit that relationship? Since the mission is a multi-agent coverage problem, we also need to ensure that the proposed coverage algorithms are scalable with the number of robots. We address these two questions in the following sections.

\section{Information Assimilation}
\label{sec:info_assimilation}

In this section, we discuss how the \ac{GP} model (\Cref{assumption:separable}) determines two functions: (A)~the information decay rate at each $p \in \Dcal$, and (B)~the information gain rate at each $p \in \Dcal$ due to measurements taken from a robot's position $r_i = \Phi(x_i) \in \Dcal$. We consider the hyperparameters of the \ac{GP} to be specified and constant, although some strategies for estimating these are discussed in the simulation section.

\subsection{\acl{KF} Model}

First, we show that the \ac{KF} is the optimal state estimator to estimate the spatiotemporal field $f$. As shown in \Cref{section:sde}, the process model for $f$ sampled at $N_G$ grid points is a linear stochastic differential equation with state $\bm s \in \R^{n_k N_G}$. We now show that the measurements~\eqref{eqn:robot_measurements} are a linear function of $\bm s$.\footnote{In~\cite{carron2016machine}, the measurements must be taken at one of the grid points. Here we extend the result to allow measurements at non-grid points.}

Consider $N_R$ robots at positions $P_R = \{ \Phi(x_{i}) \}_{i=1}^{N_r}$ where each robot makes a measurements $y_{k, i}$ as in~\eqref{eqn:robot_measurements}. However, the state $\bm s$ corresponds to the grid points $P_G$, not necessarily coinciding with the measurement locations $P_R$. To account for this, we use spatial correlation based on the Gaussian Process model for $f$: 
\eqn{
\bmat{ \bm f(t_k)\\ \bm y_k} \sim \Ncal \left( 0 , \bmat{ 
\K_{GG} & \K_{GR}\\
\K_{RG} & \K_{RR} + \sigma_m^2 I
}
\right)
}
where $\K_{GG}, \K_{GR}, \K_{RG}, \K_{RR}$ are the kernel matrices for the sets of points $P_G, P_R$, and $\bm y_k = \bmat{ y_{k, 1} & \cdots & y_{k, N_R}}^T$.

Using~\eqref{eqn:SDE_f}, $\bm y_k$ conditioned on the state $\bm s(t_k)$ is 
\begin{subequations}
\label{eqn:measurement_model}
\eqn{
\bm y_k | \bm s(t_k) &\sim \Ncal(H \bm s(t_k), V), \\
H &= \K_{RG} \K_{GG}^{-1} \sqrt{\K_{GG}} (I_{N_G} \otimes C) \\
V &= \sigma_m^2 I_{N_R} + \K_{RR} - \K_{RG} \K_{GG}^{-1} \K_{GR}.
}
\end{subequations}

Therefore, the environment's state space model is a linear (continuous time) process (recall~\eqref{eqn:SDE_Si}) with linear (discrete-time) measurements:
\begin{subequations}
\label{eqn:kf_model}
\eqn{
d \bm s &= (I_{N_G} \otimes A) \bm s dt + (I_{N_G} \otimes B) dW,\label{eqn:linear_process_model}\\
\bm y_k &= H \bm s(t_k) + v_k \label{eqn:linear_measurement_model}
}
\end{subequations}
where $W$ is a $N_G$-dimensional standard Wiener process, and $v_k \sim \Ncal(0, V)$. Notice that although each measurement has noise variance $\sigma_m^2 I$, the noise model in~\eqref{eqn:linear_measurement_model} has $V \geq \sigma_m^2 I$ accounting for the fact that measurements can be taken at non-grid points.

To summarize, we have a linear, time-invariance process model~\eqref{eqn:linear_process_model}, with a linear (but time-varying due to the changing measurement locations) measurement model~\eqref{eqn:linear_measurement_model}. Together, they satisfy the assumptions of the \ac{KF}, and therefore the \ac{KF} is the optimal estimator for this system~\cite{gelb1974applied}.

\subsection{Quantifying Information Gain and Decay Rates}
\label{section:clarity_dynamics}

Next, we wish to characterize the clarity dynamics, i.e., the rate of information gain and decay. In this section, we focus on the clarity dynamics of a single point $p \in \Dcal$ due to a measurement taken by a robot with position $r = \Phi(x) \in \Dcal$. Since we use the \ac{KF} to assimilate measurements, we use the earlier derived dynamics to estimate the rate of information gain. Reducing~\eqref{eqn:kf_model} for a single point $p$, the continuous time \ac{KF} model is 
\begin{subequations}
\eqn{
\dot s &= A s + B w,  &&w(t) \sim \Ncal(0, I),\\
y &= L s + v,  &&v(t) \sim \Ncal(0, V \Delta T)
}
\end{subequations}
 where $s \in \R^{n_k}$ is the state of the spatiotemporal process at $p$, $r = \Phi(x)$ is the robot's position, and 
\eqnN{
L = \frac{\ks(r, p)}{\sqrt{\ks(p,p)}} C, \quad V = \sigma_m^2 + \ks(r,r) - \frac{\ks(r, p)^2}{\ks(p,p)}.
}

Let the \ac{KF} state consist of $(\hat s, \Sigma)$, the mean and covariance. Then, the covariance has dynamics 
\eqn{
\dot \Sigma = A \Sigma + \Sigma A^T + BB^T - \Sigma L^T (V \Delta t)^{-1} L \Sigma.
}
Therefore, the  estimate of $f(t, p)$ is $\Ncal(\hat f, \Pi)$, where $\hat f = C \hat s$, and $\Pi = C \Sigma C^T$. Since, the clarity of a scalar Gaussian variable is $q = 1/(1 + \Pi)$, the clarity dynamics are 
\eqn{
\dot q = \frac{d q}{d \Pi} \dot \Pi = - q^2 C \dot \Sigma C^T.
}
Depending on the temporal kernel,\footnote{In particular, this holds for Matern-1/2 kernels.} this simplifies to
\eqn{
\label{eqn:S}
\dot q = \underbrace{S(x, p) (1 - q)^2}_{\text{clarity gain}} - \underbrace{D(p, q)}_{\text{clarity decay}}
}
where the first term defines the rate of clarity gain at $p$ due to measurements taken at $r = \Phi(x)$, while the second term defines the clarity decay rate.
\begin{remark}
    Eq.~\eqref{eqn:S} is one of our main results: the function $S: \Xcal \times \Dcal \to \R$ is the sensing function that quantifies the importance of a measurement taken from robot state $x \in \Xcal$ on the clarity of our estimate at a position $p \in \Dcal$. Similarly, $D: \Dcal \times \R \to \R$ defines the rate at which clarity about $f(t, p)$ decays due to the spatio\emph{temporal} nature of $f$. Notice the decay rate is uncontrolled, i.e., does not depend on the robot's state $x$.
\end{remark}

\begin{example}
    For Matern-1/2 temporal kernels, 
    \eqnN{
    S(x, p) &= \frac{1}{\Delta T} \frac{\ks(r, p)^2 }{\ks(p, p) \left(\ks(r,r)+\sigma_m^2\right) - \ks(r, p)^2 }\\
    W(p, q) &= 2 \lambda_t \left((\sigma_t^2+1)q^2 - q\right),
    }
    where $r = \Phi(x)$ is the position of a robot at state $x$. Since for isotropic spatial kernels $\ks(p,p') = \ks(\norm{p-p'})$,
\eqnN{
S(d) \propto \frac{\ks(d)^2}{ \ks(0)^2 + \sigma_m^2 \ks(0) - \ks(d)^2}
}
where $d = \norm{\Phi(x) - p}$ is the distance at which the measurement is taken. When $d \mapsto \ks(d)$ is nonincreasing, e.g. in the Matern and Squared Exponential kernels, $S(x, p)$ is maximized at $\Phi(x) = p$, implying that the rate of increase in clarity about $p$ is maximized when the robot is also at position $p$. This is not, in general, true, since for example in periodic or polynomial spatial kernels, $S(x, p)$ may be maximized for some $\Phi(x) \neq p$. Furthermore notice that in the limiting case of a spatiostatic environment, $\lambda_t \to 0$, and therefore the decay rate $D(p, q) \to 0$.
\end{example}

To summarize, in this information-gathering problem the spatiotemporal information to be collected is modeled using a \ac{GP}. To define a suitable coverage algorithm, we need to quantify the value of taking a measurement at some robot state $x \in \Xcal$ on the clarity gain at any other position $p \in \Dcal$. This is captured by the clarity dynamics~\eqref{eqn:S}. The key functions are $S$, the sensing function, and $W$, the decay function. Notice only $S(x, p)$ is controllable since it is the only term in~\eqref{eqn:S} that depends on the robot's state $x$. 



\section{Coverage Controllers}
\label{sec:coverage_controllers}

In this section, we use the sensitivity and decay functions in~\eqref{eqn:S} to derive two coverage controllers. The direct method chooses a control input that maximizes the rate of increase in the total clarity integrated over the domain $\Dcal$.  The indirect method determines the time that the robot should spend at each position in the domain to achieve a target clarity and then uses ergodic control to compute the control input. 
 
\subsection{ Direct Method }

The direct method minimizes the cost function 
\eqn{
J(t) = \norm{ \overline q(\cdot) - q(t, \cdot)}_{H}^{2}, 
}
a function-space norm over $p \in \Dcal$ between the current clarity distribution $q(t, p)$ and the target clarity distribution $\overline q(p)$. We use the Sobolev norm, defined in~\eqref{eqn:sobolev}, and discussed below. 

Notice that $J(t)$ does not explicitly depend on the robot's state or control input. As such, we choose to minimize $J$ over a short horizon $\delta>0$ in the future:
\eqn{
J(t + \delta) \approx J(t) + \dot J(t, x) \delta^2 + \frac{1}{2} \ddot J(t, x, u) \delta^2 + \cdots
}
where the dependency on $u$ first shows up in the $\ddot J$ term. This high-relative degree behavior is a consequence of the fact that the clarity dynamics~\eqref{eqn:S} depend on  $x$, not $\dot x$. Therefore the second derivative of $J$ must be taken for the control input to appear in the expressions. This behavior is commonly observed in the literature on coverage control, as in~\cite[Ch. 2]{bentz2020dynamic} and in~\cite{mathew2011metrics}.
Then, given control inputs $ u \in \Ucal \subset \R^m$, the controller will be of the form
\eqn{
\label{eqn:general_direct_method}
\pi(t, x) = \argmin{u \in \Ucal} \ \ddot J(t, x, u).
}
We will derive a closed-form solution for this controller. Before doing so, we justify our choices for the cost function and the control strategy.

We use the Sobolev norm for the following reasons. In~\cite{bentz2020dynamic}, a differentiable sensing functional (an analog of $S$) is used with the generalized transport theorem to compute an analog of $\ddot J(t, x, u)$. However, this approach often leads to local minima, where $\ddot J(t, x, u)$ becomes independent of $u$. This happens when all of the local information has been collected, and there is no preference for the controller to move in one direction over the other. To address this, \cite{bentz2020dynamic} proposed combining the local search strategy with a global strategy, where the controller would choose a new global waypoint when the local controller reaches a local minimum. In our work, we use the Sobolev space norm instead of the $\ell_2$ norm, and this allows the controllers to have a multispectral property~\cite{mathew2011metrics} - it prioritizes global coverage before prioritizing local coverage. 

Second, to evaluate $\ddot J$, we use the clarity dynamics we derived in~\Cref{section:clarity_dynamics}. This is in contrast to earlier works that used heuristic expressions to quantify coverage, and coverage dynamics~\cite{haydon2021dynamic, bentz2020dynamic}. As such, the derived controllers depend explicitly on the spatiotemporal field's kernel, and the sensing capabilities (in particular the sampling period $\Delta T$ and measurement noise $\sigma_m$) of the robots.


Next, we derive the controller. The cost function is
\eqn{
J(t) &= \norm{ \overline q(\cdot) - q(t, \cdot)}_{H}^{2} = \sum_{ l \in \naturals^d} \Lambda_l \left( \hat{\overline q}_l - \hat{q}_l(t) \right)^2,
}
where $\hat{\overline{q}}_l = \langle \overline{q}, b_l \rangle$, $\hat{q}_l(t) = \inner{q_l(t, \cdot), b_l}$ are the inner products of $\overline q(\cdot)$ and $q(t, \cdot)$ with the $l$-th basis function of the \ac{DCT}. Recall the notation $\langle a, b_l \rangle$, and $\Lambda_l$ was defined in~\Cref{sec:ergodic}. After some algebraic calculations, one can show that the first and second time-derivatives of $J$ are:
\eqnN{
\dot J(t, x) &= \sum_{l \in \naturals^d} -2 \Lambda_l ( \hat{\overline{q}}_l - \hat{q}_l(t) ) \dot{\hat{q}}_l(t, x)\\
\ddot J(t, x, u) &=  \sum_{l \in \naturals^d} 2 \Lambda_l \left ( \dot{\hat{q}}_l^2(t, x) - ( \hat{\overline{q}}_l - \hat{q}_l(t) ) \ddot{\hat{q}}_l(t,x, u)  \right ) 
}
where $\dot{\hat{q}}_l(t, x), \ddot{\hat{q}}_l(t, x, u)$ are
\eqnN{
\dot{\hat{q}}_l &= \frac{d}{dt} \inner{ q(t, \cdot), b_l}\\
&= \int_{p \in \Dcal} \left(S(x, p)(1-q(t, p))^2 - W(p, q(t, p))\right) b_l(p) dp
}
where $S$ is as defined in~\eqref{eqn:S}. Similarly,
\eqnN{
\ddot{\hat q}_l &= \frac{d^2}{dt^2}\int_{p \in \Dcal} q(t, p) b_l(p) dp = \hat B_l(t, x) \dot x + \Ocal,
}
where $\Ocal$ collects terms independent of $\dot x$ (and therefore $u$), and $\hat B_l(t, x) \in \R^{1 \times n}$ is as defined as
\eqn{
\hat B_l(t, x) &= \left \langle (1-q(t, \cdot))^2  \frac{\partial S}{\partial x}(x, \cdot), b_l\right \rangle \label{eqn:BhatL}.
}
Therefore, we have 
\eqnN{
\ddot J(t, x, u) &= \sum_{l \in \naturals^d} - \Lambda_l ( \hat{ \overline{q}}_l - \hat q_l(t)) \hat B_l(t, x) \dot x + \Ocal\\
&= - L(t, x) (F(x) + G(x) u) + \Ocal
}
where we define
\eqnN{
L(t, x) = \sum_{l \in \naturals^d} \Lambda_l ( \hat{ \overline{q}}_l - \hat q_l(t)) \hat{B}_l(t, x).
}
Therefore, the choice of $u$ that minimizes $J(t + \delta)$ yields a feedback controller $\pi_{dir}: \R \times \Dcal \to \Ucal$,  
\eqnN{
\pi_{dir}(t, x) = \argmin{u \in \Ucal}  \ -L(t, x) G(x) u 
}
If $\Ucal = \{ u \in \R^m : \norm{u} \leq u_{max} \}$, and $L(t, x) G(x) \neq 0$, 
\eqn{
\pi_{dir}(t, x) = u_{max} \frac{G(x)^T L^T(t, x)}{\norm {L(t, x) G(x)}}.
}

Proving that $L(t, x) G(x) \neq 0$ for any $t, x$ is non-trivial, and will be studied in future work.

\subsection{ Indirect Method }

The second approach is inspired by ergodic control. Ergodic control uses a \ac{TSD} to determine the feedback control law, as discussed in~\Cref{sec:ergodic}. Here we derive a principled method to construct the \ac{TSD} based on the information assimilation algorithm discussed in~\Cref{sec:info_assimilation}. 

The key idea is to set the \ac{TSD} to be the time required for the clarity of our estimate of $f$ to increase from its current value to a specified target clarity $\overline q(p)$, assuming the robot was making measurements from $x = p$. To compute this, we solve the differential equation~\eqref{eqn:S} and determine $T(q, \overline q)$, i.e., the time required to increase the clarity from $q$ to $\overline q$. 

Then, given the target clarity distribution $\overline{q}: \Dcal \to [0, 1]$, and the current clarity distribution $q(t, \cdot): \Dcal \to [0, 1]$, the \ac{TSD} can be specified as follows:
\eqn{
\operatorname{TSD}(t, p) = \begin{cases}
    T(q(t, p), \overline{q}(p)) & \text { if } q(t, p) \leq \overline{q}(p)\\
    0 & \text{ else }
\end{cases}.
}
This equation has an analytic solution, see~\cite[Appendix]{agrawal2024multi}.


Finally, we can use the ergodic control method described in~\cite{mathew2011metrics} to design a feedback controller for the system, 
\eqn{
\pi_{ind}(t, x) = \pi_{ergo}(t, x, \operatorname{TSD})
}

\subsection{Extension to Multi-Robot Coverage Control}

Our proposed coverage controllers have been presented for the single-robot cases above. Here we discuss the extension and implementation of these methods in the multi-agent case, where multiple robots have to decide how to move to collect information. We assume that they can synchronize their information by sharing the clarity map, $q(t, p) \ \forall p \in \Dcal$, over a centralized setting, i.e., that they are connected over a complete graph so that each robot has access to a centrally stored clarity map. The extensions to distributed settings are left for future work. 

Notice that both proposed controllers are feedback controllers, depending on the robot's position, and the clarity map $q(t, p)$. Therefore, the control input for each agent can be computed as $u_i = \pi(t, x_i, q)$, where $x_i$ denotes the position of the $i$-th agent, and $\pi \in \{ \pi_{dir}, \pi_{ind} \}$ can be either control strategy. In the indirect approach, we must also share the history of positions visited by the agents.

As the robots move using the coverage controllers, the robots make measurements of the spatiotemporal field from their respective positions. These measurements are assimilated into a single estimate of the spatiotemporal field using the \ac{KF} model. The information assimilation is currently performed centrally, although future work will look into distributed methods of maintaining the estimate.


\section{Simulations}

\begin{figure}
    \centering
    \includegraphics[width=0.96\linewidth]{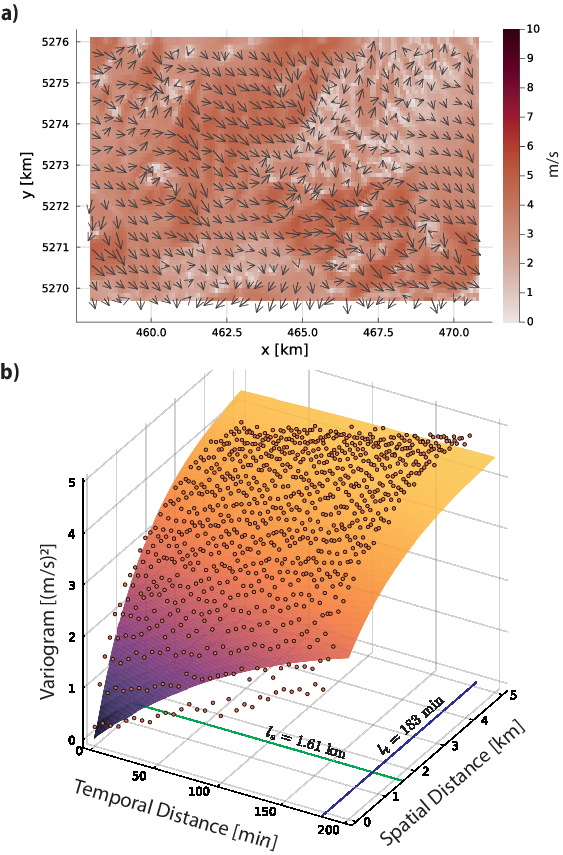}
    \caption{Wind data from WegenerNet~\cite{schlager2017generation}. (a) Wind speed and direction on Jan 1, 2023, 00:00, (b) Variogram showing the spatiotemporal correlation of the data. Surface shows the fitted kernel.}
    \label{fig:windData}
\end{figure}

\begin{figure*}
    \centering
    \includegraphics[width=0.99\linewidth]{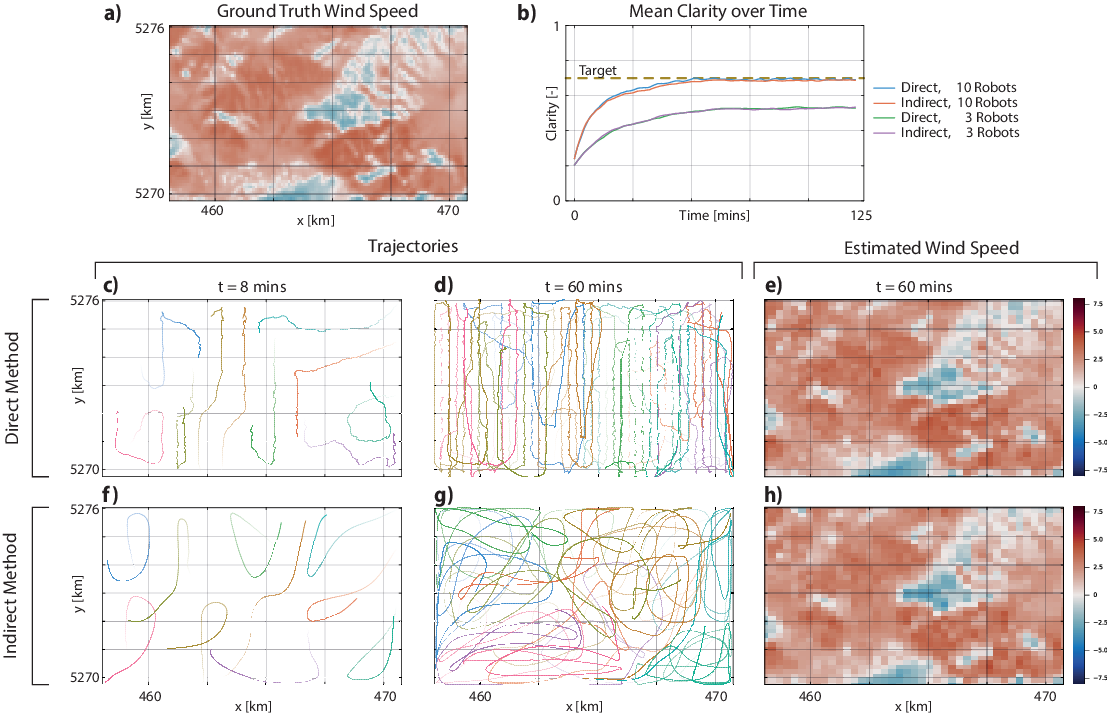}
    \caption{Simulation results. (a)~shows the ground truth wind speed at the end of the simulation. (b)~shows the mean clarity against time. The mean is taken spatially. (c-e)~show the behavior of the direct method. (f-h)~show the behavior of the indirect method. (c, d, f, g) show the trajectories of the ten robots after eight minutes and after sixty minutes.  (e, h)~show the estimated wind speed, and it closely matches the ground truth in (a). }
    \label{fig:trajs}
\end{figure*}

In this section, we report the simulation results of an information-gathering mission. As a prototypical example, we consider the collection of wind data using a team of ten aerial robots. The robots perform a two-hour mission, and we aim to maximize the clarity of the wind field over the domain by the end of the mission. Our evaluation metric is both the accuracy of the reconstruction, as well as the average clarity over the mission domain.

The mission domain is a $12.7 \times 6.3$~km$^2$ region of southeastern Austria, located near 46.93$^\circ$~N, 15.90$^\circ$~E, chosen because of a high-quality ground-truth data set available from WegenerNet~\cite{schlager2017generation}. The dataset provides wind speeds over the domain at a resolution of 100~m and 30 minutes. The mission domain is particularly challenging due to its high weather and climate variability~\cite{schlager2017generation}. Over the domain considered, the maximum wind speed is 13~m/s.

Each robot is capable of measuring the local x- and y-wind speed every 5~seconds. Each measurement is perturbed by noise with $\sigma_{m} = 0.5$~m/s. The robots are modeled as single-integrators with a maximum speed of 15~m/s. We use the \ac{KF} model with a spatial grid resolution of 320 and 160~m in the x- and y-directions to model the state of the environment. 

The spatial and temporal hyperparameters were estimated using techniques from geostatistics~\cite{wackernagel2003multivariate, christianson2023traditional}. In particular, we constructed a variogram of the dataset and used a least-squares fit to both the Matern-1/2 and the Squared Exponential kernels. The Matern-1/2 kernel fits the data better and is depicted in~\Cref{fig:windData}b. The resulting kernel is of the form $k(t, p, t', p') = \sigma^2 \exp{(-\abs{t-t'} / l_t)}\exp{(- \norm{x - x'} / l_s)}$, where $\sigma = 2.11$~m/s, $l_t = 183$~min, $l_s = 1.61$~km. Fitting the kernel using the variogram was computationally much faster and more reliable than the nonlinear minimization of the log-likelihood method of~\cite{williams2006gaussian}. See~\cite[Appendix]{agrawal2024multi} for additional details.

Simulations were run using both the direct and the indirect control strategies, and the results are summarized in~\Cref{fig:trajs}. Fig.~\ref{fig:trajs}(a) shows the ground-truth data to be estimated.\footnote{In the interest of space, only the x-component of the speed is shown. Refer to~\cite{repo} for additional figures.}

Fig.~\ref{fig:trajs}(b) shows the change in average clarity over time as the robots explore the environment. Both the direct and indirect methods result in an almost identical average clarity at each timestep. 
Furthermore, after about an hour of exploration, the average clarity reaches a steady state value. This shows that due to the information decay rate, even as the robots continually explore the environment, the average clarity cannot be increased further. 

Fig.~\ref{fig:trajs}(c,d,f,g) show the trajectories using both controllers. Figs.~\ref{fig:trajs}(c, d) show the trajectories of the direct method after 8~and 60~mins, and Figs.~\ref{fig:trajs}(f, g) show the corresponding trajectories of the indirect method. The trajectories generated by the two methods are remarkably different - in the direct method, the trajectories are jagged and tend to follow straight lines. This is because of $\partial S/\partial x$ in~\eqref{eqn:BhatL}, which places significant benefit on local data collection. In contrast, the indirect method creates smoother trajectories.

Fig.~\ref{fig:trajs}(e, h) show the estimated wind speed at $t$=60~min. Comparing these to the ground truth in Fig.~\ref{fig:trajs}(a), is it clear that both methods estimate the wind field accurately. 

In Fig.~\ref{fig:trajs}(b), we also compare the behavior when using three robots to that of using ten robots. As expected, when there are ten agents the mean clarity is higher (and increases faster) than when there are only three agents.

\section{Conclusions}

In conclusion, this paper addresses the design of cooperative multi-agent coverage controllers, where the information is shared centrally, but the control decisions are made by each robot independently. We identified a gap between information assimilation algorithms and coverage controllers. Therefore we proposed a method to quantify the value/impact that taking measurements in a domain has on the clarity of our estimate of other parts of the domain. To this end, we utilized Gaussian Processes to model the environment, as well as our earlier work on the clarity dynamics, which in effect quantifies the information gained about the domain due to measurements. We saw that the relative value of measurements is captured by a function $S$. We used this function to propose two new coverage controllers that, although qualitatively different, still cover the domain and collect information accurately. The concepts were demonstrated through a simulation study of collecting information about a wind field. 

A key limitation of this work is that we assumed the spatial and temporal hyperparameters of the Gaussian Process were fixed and known a priori. Although a method was described to obtain these hyperparameters from data, in our future work we will aim to develop an online method to estimate the hyperparameters and choose trajectories that improve the quality of the hyperparameters. 
Finally, it would also be interesting to look into methods to ensure the safety of the robots with a safety constraint that depends on the information collected online. In such a scenario, the objective of collecting information must be weighed against the importance of not violating safety constraints. 

\bibliographystyle{IEEEtran}
\bibliography{IEEEabrv,biblio}

\newpage

\newpage

\section*{Appendix}

\subsection{Marginal and Conditional Distributions}

Consider a random variable $Z \in \R^{n + m}$, given by 
\eqnN{
Z = \bmat{X\\Y} \sim \Ncal \left(\bmat{\mu_x \\ \mu_y}, \bmat{ \Sigma_{xx} & \Sigma_{xy} \\
\Sigma_{yx} & \Sigma_{yy}} \right )
}
Then the marginal distributions are given by
\eqnN{
X \sim \Ncal \left( \mu_x, \Sigma_{xx} \right)\\
Y \sim \Ncal \left( \mu_y, \Sigma_{yy} \right)
}

and the conditional distributions are given by 
\eqnN{
(X | Y = y) &\sim \Ncal (\mu, \Sigma), \\
\mu &= \mu_x + \Sigma_{xy} \Sigma_{yy}^{-1} \left( y -  \mu_y \right) \\
\Sigma &= \Sigma_{xx} - \Sigma_{xy} \Sigma_{yy}^{-1} \Sigma_{yx} 
}

Now consider two random variables $X, Y$, related by 
\eqnN{
X &\sim \Ncal(\mu, P)\\
(Y|X=x) &\sim \Ncal( C x, R )
}
where $X \in \R^n$, $Y \in \R^m$, $C \in \R^{m \times n}$, $P \in \pd^n$, $R \in \pd^m$. 

What this means is that we have an observation model 
\eqnN{
y = C x + w, \quad w \sim \Ncal(0, R)
}

\subsection{Gaussian Processes}

The kernel of a \ac{GP} is defined by the following property
\begin{definition}
    The \emph{kernel function} of a Gaussian Process $Z \sim \GP(m(x), k(x, x'))$ with mean function $m : \R^d \to \R$ and kernel function $k: \R^d \times \R^d \to \R$ is defined as
    \eqnN{
    k(x, x') = \E \left[ \left(Z(x) - m(x) \right) \left(Z(x') - m(x') \right) \right].
    }
\end{definition}

\begin{example}
The $\nu$-th order \emph{Matern kernel} is given by 
\eqnN{
k_\nu(x_1, x_2) = \sigma^2 \frac{2 ^ {1 - \nu}}{\Gamma(\nu)} \left( \sqrt{2\nu} \lambda d \right )^{\nu} K_\nu \left( \sqrt{2\nu} \lambda d \right),
}
where $\Gamma$ is the gamma function, $K_\nu$ is the modified Bessel function of the second kind, $d = \norm{ x_1 - x_2 }$, and $\sigma, \lambda > 0 $ are parameters of the kernel.
The half-integer Matern kernels are given by 
\eqnN{
k_{1/2}(x_1, x_2) &= \sigma^2 \exp {\left( - \lambda d \right)}\\
k_{3/2}(x_1, x_2) &= \sigma^2 \left(1 + \sqrt{3} \lambda d \right) \exp { \left( - \sqrt{3}\lambda d \right) }\\
k_{5/2}(x_1, x_2) &= \sigma^2 \left(1 + \sqrt{5} \lambda d  + (5/3)\lambda^2 d^2 \right) \exp {\left( - \sqrt{5} \lambda d \right)}
}
where $d = \norm{ x_1 - x_2 }$, and $\sigma, \lambda > 0$ are hyperparameters of the kernel. 
\end{example}

\subsection{Variograms}

This section establishes a method to determine the hyperparameters of a Gaussian Process using an Empirical Variogram. This method is significantly more computationally efficient and accurate than standard methods of minimizing the marginal log-likelihood but is only suitable for isotropic kernels. 

Consider the data set $D = \{ (x_i, y_i) \}_{i=1}^{N}$, where $x_i \in \R^d, y_i \in \R$. The goal is to determine the parameters of an isotropic kernel $k: \R^d \times \R^d \to \R$ that best fits the data. 

\begin{definition}
\cite[Eq 7.6]{wackernagel2003multivariate}
    The \emph{theoretical variogram} of a stationary random field $Z : \R^d \to \R$ with zero mean is 
    \eqnN{
    \gamma(d) = \frac{1}{2} \E \left[ \left(Z(x') - Z(x) \right)^2 \right],
    }
    where $\norm{x' - x} = d$. The expectation is taken over $x, x' \in \R^d$.
\end{definition}

The theoretical variogram is related to \ac{GP} kernels as follows:
\begin{lemma}
    Suppose $Z$ is a zero-mean and isotropic Gaussian Process. Then the kernel $k: \R \to \R$ and the theoretical variogram $\gamma: \R \to \R$ are related by
    \eqnN{
    \gamma(d) = k(0) - k(d)
    }
\end{lemma}
\begin{proof}
    For brevity, let $f_1 = f(x_1), f_2 = f(x_2)$.
    By the definition of the kernel, for a zero-mean GP
    \eqnN{
    k(x_1, x_2) &= E[ (f(x_1) - m(x_1)) ( f(x_2) - m(x_2))]\\
    &= E[ f_1 f_2 ]
    }
    
    Similarly, from the definition of the theoretical variogram,
    \eqnN{
    \gamma(d) &= \frac{1}{2} E[ (f(x_1) - f(x_2))^2]\\
    &= \frac{1}{2} E [f_1^2] - E[f_1 f_2] + \frac{1}{2} E[f_2^2]\\
    &= \frac{1}{2} ( E[f_1^2] + E[f_2^2] ) - E[f_1 f_2]\\
    &= \frac{1}{2}( k(x_1, x_1)  + k(x_2, x_2) ) - k(x_1, x_2)\\
    &= \frac{1}{2} (2 k(0)) - k(d)
    }
    using $d = x_2 - x_1$. 
    Therefore, 
    \eqnN{
    \gamma(d) = k(0) - k(d).
    }
\end{proof}

\begin{corollary}
In the spatiotemporal case, if the kernel is 
\eqn{
k(t, x, t', x') = k_t(t, t') k_s(x, x')
}
the theoretical variogram is 
\eqn{
\gamma(d_t, d_s) = k_t(0)k_s(0) - k_t(d_t)k_s(d_s)
}
\end{corollary}

We can use this Lemma to determine the parameters of the kernel. In particular, consider the empirical variogram:
\begin{definition}
    The \emph{empirical semi-variogram} given data $D$ is $\gamma: \R \to \R$, 
    \eqnN{
    \gamma(d) = \frac{1}{2|N(d)|} \sum_{N(d)} (y_i - y_j)^2
    }
    where $N(d) \subset \integers \times \integers$ is the set of pairs $(i, j)$ such that $\norm{x_i - x_j} \in (d - \epsilon, d+\epsilon)$ for some $\epsilon > 0$.
\end{definition}

Then, given data $D$, we construct the empirical variogram. For a given kernel $k$ with hyperparameters $\theta$, we can compute the corresponding theoretical variogram, and use least-squares fitting to determine the set of hyperparameters $\theta$ that best fit the data $D$.




\subsection{Gaussian Processes to Stochastic Differential Equations}

This section explains the equivalence between \ac{GP} and \ac{SDE} for a class of kernel functions. In this section, we focus on scalar \acp{GP} with zero mean, 
\eqnN{
f(t) \sim \GP(0, k(t, t')),
}
where $k : \R \times \R \to \R$ is denoted with $t$ to remind the reader that we consider a single (i.e. temporal) dimension. 

We use the following convention of a Fourier Transform\footnote{In \texttt{Mathematica}, one must specify \texttt{FourierParameters -> \{1, -1\}} to yield the correct convention.} of a function $g: \R \to \R$:
\begin{definition}
    The Fourier Transform of a function $g: \R \to \R$ is the function $G: \R \to \R$, 
    \eqnN{
    G(\omega) = \Fcal[g](\omega) = \int_{-\infty}^{\infty} g(t) e^{-i \omega t} dt
    }
    The \emph{Inverse Fourier Transform} is 
    \eqnN{
    g(t) = \Fcal^{-1}[G](t) = \frac{1}{2\pi} \int_{-\infty}^{\infty} G(\omega) e^{i\omega t} d\omega
    }
\end{definition}

This convention has the following properties:
\eqnN{
\Fcal \left[ \frac{d^n g}{dt^n}\right](\omega) = (i\omega)^n \Fcal[g](\omega)
}

The Wiener-Khinchin theorem relates the kernel function to the power spectrum of a stochastic process:
\eqnN{
S(\omega) = \Fcal[k](\omega)
}
here, we write $k(\tau) = k(t, t')$ for any $\abs{t-t'} = \tau$. When $S$ is a rational function of even order $2n_k$, we can decompose $S$ as 
\eqnN{
S(\omega) = L(\omega) L(-\omega)
}
where 
\eqnN{
L(\omega) = \frac{ b_{n_k-1} (i \omega)^{n_k-1} + b_{n_k-2}(i\omega)^{n_k-2} + \cdots + b_0}{
(i\omega)^{n_k} + a_{n_k-1}(i\omega)^{n_k-1} + \cdots + a_0}
}

Given this decomposition, we know that the stochastic process $f$ is a realization of a white noise process $W(t)$ that has been colored using the transfer function $L(\omega)$. Therefore, the state-space model of the system is 
\eqnN{
ds &= A s dt + B dW\\
z &= C s
}
where $s \in \R^{n_k}$ is the state, $W(t)$ is the standard (1D) white noise process. The output $z(t)$ will have the correct kernel function. Here, the constant matrices $A \in \R^{n_k \times n_k}$, $B \in \R^{n_k \times 1}$, $C \in \R^{1 \times n_k}$ are 
\eqnN{
A &= \bmat{ 
0 & 1 & 0 & \cdots & 0\\
0 & 0 & 1 & \cdots & 0\\
  &   &   & \ddots &  \\
0 & 0 & 0 & \cdots & 1\\
-a_0 & -a_1 & -a_2 & \cdots & -a_{n_k-1}
}, \quad B = \bmat{ 0 \\ 0 \\ \vdots \\ 0 \\ 1}\\
C &= \bmat{ b_0 & b_1 & b_2 & \cdots & b_{n_k-1}}
}

To create a realization of $f$ that has the desired kernel function, simulate the \ac{SDE} starting from $s_0 \sim N(0,\Sigma_0)$, where $\Sigma_0 \in \R^{n_k \times n_k}$ is the solution to the Lyapunov equation $A X + X A^T + BB^T  = 0$.

Finally, the discrete time version of this, with a sampling period $\Delta t$ is
\eqnN{
s_{k+1} &= \Phi s_k + w_k, \quad w_k \sim \Ncal(0, W)\\
z_k &= C s_k
}
where 
\eqnN{
\Phi &= e^{A \Delta t}\\
W &= \int_{0}^{\Delta t} e^{A \tau} B B^T e^{A^T \tau} d\tau
}

Some analytic expressions are derived below. 
\begin{example}[Matern 1/2]
The 1D Matern-1/2 kernel is 
\eqnN{
k_{1/2}(d) &= \sigma^2 \exp {\left( - \lambda d \right)}
}
It has a power-spectral density
\eqnN{
S_{1/2}(\omega) &=  \frac{2  \lambda \sigma ^2}{\lambda^2 + \omega ^2} 
}
and rational decomposition
\eqnN{
L_{1/2}(\omega) &= \frac{  \sigma \sqrt{2 \lambda} }{(i \omega) + \lambda}
}
Therefore, the state-space representation is 
\eqnN{
\left(\begin{array}{c|c}
A & B \\
\hline
C 
\end{array}\right) = 
\left( 
\begin{array}{c | c}
-\lambda & 1\\
\hline 
\sigma \sqrt{2 \lambda}
\end{array}\right)
}
\end{example}

\begin{example}[Matern 3/2]
The 1D Matern-3/2 kernel is 
\eqnN{
k_{3/2}(x_1, x_2) &= \sigma^2 \left(1 + \sqrt{3} \lambda d \right) \exp { \left( - \sqrt{3}\lambda d \right) }
}
It has a power-spectral density
\eqnN{
S_{3/2}(\omega) &=  \frac{12 \sqrt{3} \lambda^3 \sigma ^2  }{\left(3 \lambda^2 + \omega ^2\right)^2}
}
and rational decomposition
\eqnN{
L_{3/2}(\omega) &=\frac{ \sqrt{12 \sqrt{3}} \lambda ^{3/2} \sigma }{  (i\omega)^2 + 2 \sqrt{3} \lambda  (i\omega) + 3\lambda^2}
}
Therefore, the state-space representation is 
\eqnN{
\left(\begin{array}{c|c}
A & B \\
\hline
C 
\end{array}\right) = 
\left( 
\begin{array}{cc | c}
0 & 1  & 0\\
-3\lambda^2 & -2 \sqrt{3} \lambda & 1\\
\hline 
\sqrt{12 \sqrt{3}} \lambda ^{3/2} \sigma & 0 
\end{array}\right)
}
\end{example}

\begin{example}[Matern 5/2]
The 1D Matern-5/2 kernel is 
\eqnN{
k_{5/2}(x_1, x_2) &= \sigma^2 \left(1 + \sqrt{5} \lambda d  + (5/3)\lambda^2 d^2 \right) \exp {\left( - \sqrt{5} \lambda d \right)}
}
It has a power-spectral density
\eqnN{
S_{5/2}(\omega) &=  \sigma ^2 \frac{400 \sqrt{5} \lambda^5 }{3 \left(5\lambda^2 + \omega ^2\right)^3} 
}
and rational decomposition
\eqnN{
L_{5/2}(\omega) &= \frac{ \sqrt{\frac{400 \sqrt{5}}{3}} \lambda^{5/2} \sigma }{
(i\omega)^3 +  3 \sqrt{5} \lambda (i\omega)^2 +  5 \sqrt{5} \lambda^3}
}
Therefore, the state-space representation is 
\eqnN{
\left(\begin{array}{c|c}
A & B \\
\hline
C 
\end{array}\right) = 
\left( 
\begin{array}{ccc | c}
0 & 1 & 0    & 0 \\
0 & 0 & 1    & 0\\
-5 \sqrt{5}\lambda^3 & -15 \lambda^2 & -3 \sqrt{5} \lambda     & 1\\
\hline 
\sqrt{\frac{400 \sqrt{5}}{3}} \lambda^{5/2} \sigma & 0 & 0
\end{array}\right)
}
\end{example}

\subsection{Solutions to the Ricatti Equation}

\begin{lemma}
    Consider a differential equation
    \begin{subequations}
        \label{eqn:standard_ricatti}
    \eqn{
    y'(t) &= -\alpha y(t)^2 - \beta y(t) - \gamma\\
    y(0) &= y_0
    }
    \end{subequations}
    where $\alpha, \beta, \gamma \in \R$, $\alpha \neq 0$, and $\delta^2 = \beta^2 - 4 \alpha \gamma  > 0$. 
    Then, the solution is given by
    \eqn{
    y(t) = \frac{1}{2\alpha} \left(-\beta +\delta +\frac{2 \delta  \rho _0}{\left(2 \delta +\rho _0\right) e^{\delta  t}-\rho _0}\right)
    \label{eqn:standard_ricatti_solution}
    }
    where $\rho_0 = \beta -\delta +2 \alpha  y_0$.
\end{lemma}
\begin{proof}
    This is a second-order nonlinear differential equation, also known as the scalar Ricatti equation. As proposed in~\cite[Ch. 2.15]{ince1927ordinary}, consider the substitution
    \eqn{
    y(t) = \frac{u'(t) }{\alpha u(t)}
    }
    Then, it is equivalent to the following differential equation 
    \begin{subequations}
    \eqn{
    u''(t)  &= -\beta u'(t) - \alpha \gamma u(t)\\
        y_0 &= \frac{u'(0)}{\alpha u(0)}
    }
    \end{subequations}
    This linear second-order differential equation has a unique solution
    \eqn{
    u(t) = \left(c_2 e^{\delta  t}+c_1\right) e^{-\frac{1}{2} t (\beta +\delta )}
    }
    where $\delta = \sqrt{\beta - 4 \alpha \gamma}$ and the constants $c_1,c_2$ depend on the boundary condition. Evaluating the boundary conditions, we have the relationship
    \eqn{
    y_0=\frac{\frac{1}{2} (c_1+c_2) (-\beta -\delta )+c_2 \delta }{\alpha  (c_1+c_2)}
    }
    Evaluating $y = u'/(\alpha u)$, we have 
    \eqn{
    y(t) = \frac{1}{2\alpha} \left(-\beta +\delta -\frac{2 c_1 \delta }{c_2 e^{\delta  t}+c_1} \right)
    }
    Plugging in the boundary condition, we arrive at
    \eqn{
    y(t) = \frac{1}{2\alpha} \left(-\beta +\delta +\frac{2 \delta  \rho _0}{\left(2 \delta +\rho _0\right) e^{\delta  t}-\rho _0}\right)
    }
    where $\rho_0 = \beta -\delta +2 \alpha  y_0$, independent of $c_1, c_2$. 
\end{proof}

\begin{corollary}
    The limiting value of \eqref{eqn:standard_ricatti_solution} is 
    \eqn{
    y_\infty = \lim_{t \to \infty} y(t) = \frac{\delta -\beta}{2 \alpha}.
    }
\end{corollary}

\begin{corollary}
    The inverse of \eqref{eqn:standard_ricatti_solution} is given by 
    \eqn{
    t = \frac{1}{\delta} \log \left(\frac{\rho_0 \left(2 \delta +\rho_f\right)}{\rho_f \left(2 \delta +\rho_0\right)}\right)
    }
    where $\delta^2 = \beta^2 - 4 \alpha \gamma$, $\rho_0 = \beta -\delta +2 \alpha  y_0$ and $\rho_f = \beta -\delta +2 \alpha y_f$.
\end{corollary}

\end{document}